\documentclass[aps,superscriptaddress,twocolumn,nofootinbib,preprintnumbers]{revtex4}

\usepackage{amsmath}
\usepackage{graphicx,tabularx}
 
\newcommand{\sla}[1]{/\!\!\!#1}

\newcommand{\go}   {\tilde{g}}
\newcommand{\st}[1]{\tilde{t}_{{#1}}}
\newcommand{\sbx}[1]{\tilde{b}_{{#1}}}
\newcommand{\sq}[1]{\tilde{q}_{{#1}}}

\newcommand{\se}[1]{\tilde{\ell}_{{#1}}}
\newcommand{\stau}[1]{\tilde{\tau}_{{#1}}}
\newcommand{\nn}[1]{\tilde{\chi}^0_{{#1}}}

\newcommand{\cp}[1]{\tilde{\chi}^+_{{#1}}}
 
\providecommand{\SP}{\scriptscriptstyle}
\newcommand{\mgo}   {m_{\tilde{g}}}

\newcommand{\msb}[1]{m_{\tilde{b}_{\SP {#1}}}}

\newcommand{\msl}[1]{m_{\tilde{\ell}_{\SP {#1}}}}
\newcommand{\mstau}[1]{m_{\tilde{\tau}_{\SP {#1}}}}
\newcommand{\mnn}[1]{m_{\tilde{\chi}^{\SP 0}_{\SP {#1}}}}

\def\ie{{\it i.e.\;}}
\def\eg{{\it e.g.\;}}
\def\cf{{\it c.f.\;}}

\newcommand{\gev}{~{\ensuremath\rm GeV}}
\newcommand{\tev}{~{\ensuremath\rm TeV}}
\newcommand{\fb}{~{\ensuremath\rm fb}}

\newcommand{\pb}{~{\ensuremath\rm pb}}

\newcommand{\ifb}{~{\ensuremath\rm fb^{-1}}}
 
\begin{document}
 
\date{\today}
 
\title{It's a Gluino!}

\preprint{MPP-2006-42}

\author{Alexandre Alves}
\affiliation{Instituto de F\'\i sica, Universidade de S\~{a}o Paulo,
             S\~{a}o Paulo, Brazil}
\author{Oscar \'Eboli}
\affiliation{Instituto de F\'\i sica, Universidade de S\~{a}o Paulo,
             S\~{a}o Paulo, Brazil}
\author{Tilman Plehn}
\affiliation{Heisenberg Fellow, Max Planck Institute for Physics,
             Munich, Germany \\
             and School of Physics, University of Edinburgh,
             Scotland}

\begin{abstract}
  For a long time it has been known that the like--sign dilepton signature can
  help establish the existence of a gluino at the LHC.  To unambiguously claim
  that we see a strongly interacting Majorana fermion --- which we could call
  a gluino --- we need to prove that it 
  is indeed a fermion. We propose how to extract this information from 
  a different gluino decay cascade which is also used to measure its mass.
  Looking only at angular
  correlations we distinguish a universal extra dimensional
  interpretation assuming a bosonic heavy gluon from supersymmetry with a 
  fermionic gluino.  Assuming a supersymmetric interpretation, we show how the
  same angular correlations can be used to study the left--right nature of the
  sfermions appearing in the decay chain.
\end{abstract}
 
\maketitle
 
\subsection{Introduction}
\label{sec:intro}

Signals for new physics at hadron colliders largely rely on the
production and decay of strongly interacting new particles, \eg in the
case of supersymmetry~\cite{susy} the production of squarks and
gluinos~\cite{deq,prospino} with their subsequent decays. Based on
this simple fact, it is obvious how to inclusively search for these
particles: if the lightest supersymmetric partner is neutral and
stable, squarks and gluinos have to decay to (at least) one or two
jets and missing transverse momentum $\sla{p}_T$.\smallskip

Once we require one charged lepton in the squark or gluino decay we
can start testing properties of SUSY--QCD: like--sign dileptons can
for example be produced in quark--quark scattering $qq \to \sq{}
\sq{}$ via a $t$--channel gluino. This process requires
fermion--number violating interactions of the gluino, \ie it is a sign
for the Majorana nature of the $t$--channel fermion. Like--sign
dileptons also appear in gluino pair production $q\bar{q}/gg \to
\go\go$, when the Majorana gluino decays to $q \sq{}^*$ or $\bar{q}
\sq{}$ and the squark/antisquark decay yields one definite--charge
lepton~\cite{likesign}. The advantage of this SUSY-QCD signature is
that the signal process is strongly interacting, while the
(non-misidentification) backgrounds are multiple $W$ and $Z$ boson
production, \ie weakly interacting or multi--top induced. At the LHC
pairs of $500 \gev$ gluinos are copiously produced, with
cross sections of ${\cal O}(50 \pb)$ (not counting the large
associated $\sq{}\go$ production channel)~\cite{prospino}. Therefore, there is
little doubt that we will be able to extract this like--sign dilepton
signature even if its branching ratio is small.\smallskip

Motivated by electroweak baryogenesis and its requirement for light stops,
there is a variation of this like--sign dilepton signature~\cite{sabine},
namely the decay $\go \to t \st{1}^*/\bar{t} \st{1}$~\cite{decay_st}.  Because
the stop decays $\st{} \to b \cp{1} \to b W^+ \nn{1}$ and $\st{} \to t \nn{1}$
are irreducible from a top decay, the like--sign dileptons gluino events will
look just like Standard Model $t\bar{t}t\bar{t}$ production, except with an
increased rate and possibly different angular correlations.\smallskip

This recipe for using the like--sign dilepton signature to show that
new physics at the LHC incorporates a strongly interacting Majorana
fermion 
and is, therefore, likely to be SUSY--QCD unfortunately has a
loop hole. If the particle responsible for a gluino--like cascade
decay is a boson~\cite{ued, early} with an adjoint color charge, the
like--sign dilepton signature will naturally occur: two such bosons
will each decay into either a `squark--antisquark' pair or even into a
simple Standard Model $t\bar{t}$ pair and thus produce like--sign
dileptons.\bigskip

To close this loop hole we need to show that the strongly interacting particle
responsible for our like--sign dilepton events is indeed a fermion. Depending
on the supersymmetric mass spectrum, the gluino mass can be precisely
determined in the 
(not like--sign dilepton) 
cascade decay $\go \to b \sbx{1}^*/\bar{b} \sbx{1}$, where
the light sbottom decays through the long chain $\sbx{1} \to \nn{2} \to \se{}
\to \nn{1}$~\cite{edges,mgl}. The two decay chains would then 
have to be linked by comparing their detailed decay kinematics.
In the similar case of a $\sq{L}$ decay we know
how to show that the starting point of that cascade is indeed a
scalar~\cite{barr,smillie}.  To do so, the strategy includes a few crucial
steps: first, we assume (and for gluino decays with bottom tags we know) that
all outgoing Standard Model particles in the cascade decay are fermions.  In
other words, the intermediate particles in the cascade have to alternate
between fermions and bosons. To determine the spin nature of the heavy
SUSY--QCD particle all we have to do is compare the SUSY cascade with another
scenario where the new intermediate states have the same spin as the Standard
Model particles instead. Such a model are Universal Extra Dimensions (UED)~\cite{ued}
where each Standard Model particle has a heavy Kaluza--Klein (KK) partner
which can mimic the SUSY cascade decay, provided the mass spectra which
can be extracted from the decay kinematics match~\cite{early}.\smallskip

There are many observables which we can use to discriminate `typical'
UED and SUSY models, like the production rate~\cite{smillie}, ratios of branching fractions
or the mass spectrum.
At the LHC we measure only production cross sections
times branching ratios times efficiencies with fairly large errors.
In particular in the supersymmetric squark sector rate information
can be diluted through the existence of several strongly interacting
scalars with similar decays. 
Moreover, the UED as well as the SUSY mass spectra
might well to be what we currently consider `typical'. On the other
hand, direct spin information is generally extracted from angular
correlations. This kinematic information should at the end be combined 
with rate information.
Because these two approaches are independent of each other
we base our analysis exclusively on
distributions of the outgoing Standard Model fermions as predicted by
UED and by SUSY for a well--established
decay chain. All masses in the decay cascade we assume to be
measured from the kinematic endpoints of the same set of
distributions.  Because angles are not Lorentz invariants, it is much
easier to interpret invariant masses like $m_{\ell q},
m_{\ell\ell}$~\cite{barr}. Boosting the laboratory frame into the rest
frame of for example the $\nn{2}$ we can (on the generator level)
translate invariant masses and angles into each other. The only angles
which are independent of the unknown over-all event boost in the beam
direction are azimuthal opening angles, \eg between the two bottom
jets $\phi_{bb}$, whose distinguishing power we will discuss in a
separate section.\smallskip

If we knew which of the leptons in the $\sq{L}$ decay chain is the one
radiated right after the quark (the near lepton), \ie if we could link
$\ell^{+/-}$ and $\ell^{\rm near/far}$~\cite{edges,barr,smillie}, we
could simply compare $m_{q\ell}$ mass distributions to distinguish UED
from SUSY cascades.  In practice, we have to find a way to not
symmetrize over $\ell^+$ and $\ell^-$ or $q$ and $\bar{q}$. The trick
used for the $\sq{L}$ cascade is to rely on the fact that squarks are
largely produced in association with a gluino ($pp \to \sq{L} \go$),
and that squark cascade decays will preferably produce $q$ and not
$\bar{q}$ jets, even though on an event-by-event basis we cannot
distinguish the two~\cite{smillie}.\smallskip

Because of the like--sign dilepton argument described above, we are
much more interested in the spin of the gluino than in the spin of a
squark. Luckily, for the determination of the gluino spin we can
almost completely follow the squark spin argument, with the exception
of the last trick --- a Majorana gluino will always average over $q$
and $\bar{q}$, or in the case of the bottom cascade (which we can use to
measure the gluino mass) over $b$ and $\bar{b}$.  However, tagged bottom jets
require a lepton, so we can distinguish $b$ and $\bar{b}$ on an
event-by-event basis and do not have to rely on any argument linked to
the gluino production mechanism.\bigskip

The determination of quantum numbers of new particles is a necessary
addition to recent progress in determining Lagrangian mass parameters
from LHC (and ILC) measurements. We know that at the LHC we will be
able to identify new physics models based on mass spectra extracted
from decay kinematics~\cite{edges, sfitter, gordi}. In combination
with ILC measurements it is in principle possible to reconstruct all
mass parameters for example in the TeV--scale MSSM Lagrangian, not
only in the benchmark point SPS1a~\cite{sps}. However, all these
studies assume that we know the spin of the new particles, \ie we know
which operator in the Lagrangian we have to link with a measured mass.
The determination of the squark spin~\cite{barr,smillie} and now of
the gluino spin (plus the spins of the other intermediate particles
which are produced radiating Standard Model fermions) from decay
kinematics at the LHC adds crucial information to the reconstruction
of new physics at colliders --- even before we can start systematic
studies of particle thresholds at the ILC~\cite{tesla}.

\subsection{Universal Extra Dimensions}
\label{sec:ued}

Before we describe in some detail the UED Lagrangian we are using to
contrast the supersymmetric gluino cascade, we emphasize that this
paper is not about trying to discover a typical UED cascade at the
LHC. Instead, we will use UED as a straw man, which we set on fire to
shed light on the gluino cascade.\smallskip

The most notable difference between a typical UED cascade decay,
compared to a SUSY cascade decay, is that (unless we invoke additional
boundary conditions or include large radiative corrections) all
Kaluza--Klein excitations of the Standard Model particles are mass
degenerate. This means the outgoing fermions from cascade decays
become very soft, hard to identify and even harder to distinguish from
backgrounds. For example, the highly efficient $\sla{p}_T^{\rm min}$
cut with which we extract SUSY--QCD signals for Standard Model
backgrounds is far less effective for a typical UED scenario with a
lightest KK partner.\smallskip

For the sake of comparison we assume one extra dimension with size $R
\sim \tev^{-1}$~\cite{ued,early}. For each of the Standard Model
fields ($n=0$) we obtain a tower of discrete KK excitations with mass
$m^{(n)} = \sqrt{n^2/R^2 + (m^{(0)})^2}$, $n \geq 1$. For example, the
5-dimensional wave functions for an SU(2)--doublet fermion are of the
form:
\begin{equation}
\psi_d = \frac{1}{\sqrt{2\pi R}} \psi^{(0)}_{dL} 
       + \frac{1}{\sqrt{\pi R}} \sum_{n=1}^{\infty}
         \left( \psi^{(n)}_{dL} \cos\frac{ny}{R}
               +\psi^{(n)}_{dR} \sin\frac{ny}{R}
         \right)
\end{equation}
For SU(2) singlets the roles of the left and right handed $n$-th KK
excitations are reversed. Gauge bosons only involve the cosine term for the
$n$-th KK excitations. Just like in the MSSM, the spinors of the singlet ($q$)
and doublet ($Q$) KK--fermion mass eigenstates can be expressed in terms of
the SU(2) doublet and singlet fields $\psi_{d,s}$:
\begin{alignat}{5}
Q^{(n)} &=  \; \cos \alpha^{(n)} \psi_d^{(n)} 
             &+\;& \sin\alpha^{(n)} \psi_s^{(n)} \notag \\
q^{(n)} &=  \; \sin\alpha^{(n)}\gamma^5 \psi_d^{(n)} 
             &-&   \cos\alpha^{(n)}  \gamma^5\psi_s^{(n)}
\label{eq:eigen}
\end{alignat}
Their mixing angle $\alpha^{(n)}$ is suppressed by the Standard Model
fermion mass over the (large) KK--excitation mass plus one--loop
corrections:
\begin{equation}
\tan 2\alpha^{(n)}= \frac{m_f}{n/R +
  (\delta m^{(n)}_Q+\delta m^{(n)}_q)/2}
\label{eq:mixing1}
\end{equation}
The non--degenerate KK--mass terms $\delta m^{(n)}$ contain tree level
and loop contributions to the KK masses, including possibly large
contributions from non--universal boundary conditions.\bigskip

The neutral KK gauge fields will play the role of neutralinos in the
alternative description of the gluino cascade. Just as in the Standard Model,
there is a KK--weak mixing angle which for each $n$ rotates the interaction
eigenstates into mass eigenstates:
\begin{alignat}{5}
\gamma_\mu^{(n)} &=  & \cos \theta_{w}^{(n)} \, B_\mu^{(n)} 
                 &+\;& \sin \theta_{w}^{(n)} \, W_{3,\mu}^{(n)}
\notag \\
Z_\mu^{(n)}      &= -& \sin \theta_{w}^{(n)} \, B_\mu^{(n)}
                 &+&   \cos \theta_{w}^{(n)} \, W_{3,\mu}^{(n)}
\label{eq:mixing2}
\end{alignat}
The $n$-th KK weak mixing angle is again mass suppressed
\begin{equation}
  \tan 2\theta_w^{(n)} = \frac{v^2 \, g \, g_Y/2}{
                         ( \delta m^{(n)}_{W_3} )^2
                       - ( \delta m^{(n)}_B )^2
                       + v^2 \left( g^2-g^2_Y \right)/4} 
\label{eq:weinberg}
\end{equation} 
where $\delta m^{(n)}$ contains tree level as well as loop corrections to the
KK gauge boson masses. Generally $(\delta m^{(n)}_{W_3} )^2 - ( \delta
m^{(n)}_B )^2 \gg v^2 \left( g^2-g^2_Y \right)$~\cite{early} and the lightest
KK partner is the $B^{(1)}$, with basically no admixture from the heavy
$W_3^{(1)}$. Note that this formula ties the KK weak mixing angle to the mass
spectrum --- even when boundary terms are taken into account.\smallskip

To formulate an alternative interpretation of a gluino decay cascade at the
LHC we only need the first set of KK excitations ($n=1$). Higher excitations
might be used as another means to distinguish SUSY and UED signals at future
colliders, provided they are not too heavy~\cite{ued,early}.  The UED decay
chain we use to mimic a gluino decay is $g^{(1)} \to b^{(1)} \to Z^{(1)} \to
\ell^{(1)} \to \gamma^{(1)}$. In general, the KK partners of the Standard
Model particles do not have a mass spectrum similar to what we expect in SUSY.
For instance, $m_{g_1}= 640 \gev$, $m_{b_1}= 564 \gev$, $m_{Z_1}= 536 \gev$,
$m_{\ell_1}= 505 \gev$, and $m_{\gamma_1}= 501 \gev$ for $R=500\gev^{-1}$,
$\Lambda R=20$, $m_h=120 \gev$, and vanishing boundary terms at the cut--off
scale $\Lambda$~\cite{smillie,early}.\smallskip

The coupling of KK gluons to KK quarks and Standard Model quarks is crucial
for our analyses.  The coupling of the KK mass eigenstates $q^{(1)}$ and
$Q^{(1)}$ in Eq.~(\ref{eq:eigen}) to KK gluons $G^{(1)a}_\mu$ and SM quarks
$\psi^{(0)}$ after integration over the extra dimension is:
\begin{alignat}{5}
{\cal L}_{\rm QCD} &= ig_s T^a \notag \\
 \Big[ \;  &\bar{\psi}^{(0)}\gamma^\mu G^{(1)a}_\mu
                          \left( \cos\alpha^{(1)} P_L
                                +\sin\alpha^{(1)} P_R
                          \right) Q^{(1)} 
                  \notag \\
          -&\bar{\psi}^{(0)}\gamma^\mu G^{(1)a}_\mu
                          \left( \sin\alpha^{(1)} P_L
                                +\cos\alpha^{(1)} P_R
                          \right) q^{(1)}
\Big]
\label{eq:lagrangian1}
\end{alignat} 
This is analogous to the Yukawa interactions $\go$--$\tilde{q}$--$q$ in
SUSY-QCD. The gluon couplings illustrate the correspondence of the
left--right mixing angle in the squark sector with the singlet-doublet mixing
for UED. The mass suppression typically drives the mixing angle
$\alpha^{(1)}$ to zero except for the top quark. The complete set of Feynman
rules for the electroweak sector in terms of mass eigenstates can be found in
Ref.~\cite{Bringmann}. Here we just quote the electroweak Lagrangian relevant
for the couplings in long and short cascades
\begin{alignat}{5}
{\cal L}_{\rm ew} &= \notag \\
&ig \; \bar{\psi}^{(0)}\gamma^\mu P_L A^{3(1)}_\mu
                          \left(I_3 \cos\alpha^{(1)} Q^{(1)}
                               -I_3 \sin\alpha^{(1)} q^{(1)}\right)
                  \notag \\
          +&ig_Y \; \bar{\psi}^{(0)}\gamma^\mu B^{(1)}_\mu
                          \left( Y_s \sin\alpha^{(1)} P_R
                                +Y_d \cos\alpha^{(1)} P_L
                          \right) Q^{(1)}
                  \notag \\
          -&ig_Y \; \bar{\psi}^{(0)}\gamma^\mu B^{(1)}_\mu
                          \left( Y_s \cos\alpha^{(1)} P_R
                                +Y_d \sin\alpha^{(1)} P_L
                          \right) q^{(1)}
\label{eq:lagrangian2}
\end{alignat}
where $I_3$ and $Y$ are the usual isospin and hypercharge of the 
Standard Model fermions. $A^{3(1)}_\mu$ and $B^{(1)}_\mu$ are the
KK excitations of the neutral gauge bosons.

\subsection{One MSSM Example: SPS1a}
\label{sec:sps}

For a quantitative study we choose the (collider friendly) parameter point
SPS1a. The masses in the gluino decay cascade are $\mgo = 608 \gev$, $\msb{1}
= 517 \gev$, $\msb{2} = 547 \gev$, $\mnn{2} = 181 \gev$, $\msl{1} = 145 \gev$,
$\msl{2} = 202 \gev$, $\mstau{1} = 136 \gev$, $\mstau{2} = 208 \gev$, and
$\mnn{1} = 97 \gev$. The NLO production cross sections are $7.96 \pb$ for
$\go\go$, $8.02 \pb$ for $\sq{}\sq{}^*$, $26.6 \pb$ for $\sq{}\go$, and $7.51
\pb$ for $\sq{} \sq{}$. For the SPS1a parameter choice the lighter of the two
sbottoms is almost entirely $\sbx{1} \sim \sbx{L}$. The stau mixing pattern is
identical to the sbottoms, while the sleptons exhibit the opposite behavior
$\se{1} \sim \se{R}$.  The gluino mass can be measured at the percent level in
the cascade decay $\go \to \sbx{1} \to \nn{2} \to \se{1} \to
\nn{1}$~\cite{edges,mgl}.  The branching fraction for this decay is $0.4\%$.
The gluino branching fraction to one charged lepton, on which the like--sign
dilepton signature is based, is $0.4\%$ as well.\smallskip

\begin{figure}[t]
  \begin{center}
  \includegraphics[width=8.5cm]{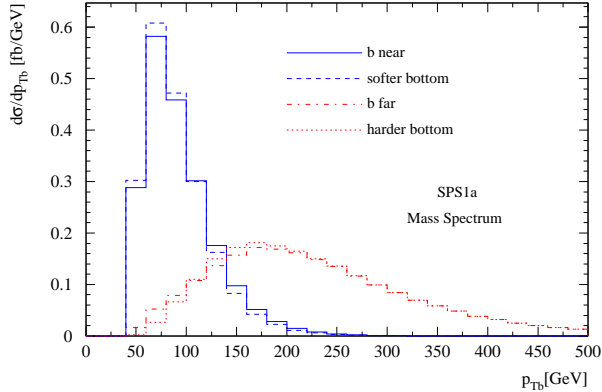}
  \end{center}
  \vspace*{-8mm}
  \caption{The transverse momentum spectra for the bottom jets in the
    gluino decay. The tagged bottom jets are ordered according to
    their appearance (near/far) in the gluon decay chain and according
    to their transverse momentum.}
\label{fig:ptb}
\end{figure}

Because the measurement of the gluino spin is most important in combination
with the observation of like--sign dileptons, we concentrate on gluino--pair
production. To avoid combinatorial backgrounds we require one gluino decay
through the short cascade with a light--flavor squark decaying into one
light--flavor jets and the LSP. If the squark is right handed and the LSP is
mostly bino, this short cascade will dominate over the long cascade which also
radiates two leptons; the gluino branching ratio through the short
squark decay chain is $41 \%$. For the second gluino we require two tagged
bottom jets (to identify the gluino--decay jets) and the long cascade through
a slepton. This selection means that it is straightforward to also include the
potentially large associated $\sq{}\go$ production, where the squark which we are not interested in decays to
a jet and the LSP.  This second production process reduces the statistical
errors significantly without having any impact on our actual gluino decay
analysis.  Our possible signal processes are
\begin{alignat}{5}
pp &\to \go\go \to jjb\bar{b}\ell^+\ell^- + \sla{p}_T \notag \\
pp &\to \sq{}\go \to jb\bar{b}\ell^+\ell^- + \sla{p}_T
\end{alignat}
where $\ell$ stands for electrons and muons, which can in principle come from
tau decays. In the following, the associated $\sq{}\go$ channel is not
included unless we explicitly state this. The dominant Standard Model
background is obviously $t\bar{t}$+jets. For the parameter point SPS1a both
$b$ jets are hard (\cf Fig.~\ref{fig:ptb}), so we do not expect any
complication identifying the gluino cascade.  If we were to extend our
analysis to the associated production with another long cascade, we could use
the mixed--flavor sample to avoid combinatorial backgrounds.\bigskip

In some scenarios, like in SPS1a, the mass hierarchy has a favorable impact on
the momentum of the jets radiated off the decay cascade. In Fig.~\ref{fig:ptb}
we see that just picking the harder of the two bottom jets we can distinguish
between `near' (gluino decay) and `far' (sbottom decay) jet on an
event--by--event basis, to construct an asymmetry.  However, for most of our
analysis we choose to ignore this spectrum dependent approach in favor of the
general method of distinguishing $b$ and $\bar{b}$ jets by the lepton charge
in the $b/\bar{b}$ tag.\medskip

The lighter of the two sbottoms and sleptons dominate the long gluino decay
chain, but in our numerical analysis we always include all scalar mass
eigenstates \ie we include intermediate $\sbx{1,2}$ as well as $\se{1,2}$ and
$\stau{1,2}$ in the cascade. True off-shell SUSY effects will be strongly
suppressed~\cite{smadgraph}.  The contribution of the heavier sbottom $\sbx{2}
\sim \sbx{R}$ to the gluino decay width is roughly five times smaller than the
$\sbx{1} \sim \sbx{L}$ contribution.  The leptonic $\tau$ decays we compute in
the collinear approximation ($m_\tau \ll p_{T, \tau}$).\smallskip

For the parton--level decay chains we include the UED spectrum in
Madgraph~\cite{madevent} and use Smadgraph~\cite{smadgraph} for the
SUSY simulation. This way we correctly treat all spin correlations.
All final--state momenta are smeared to simulate detector effects.
After including a $60\%$ $b$-tagging efficiency, $b$ and $\bar{b}$ can
be distinguished by the lepton charge in semileptonic decays ($22\%$
branching ratio times $80\%$ lepton detection efficiency) with a
$30\%$ mistag probability~\cite{btags}.  When the tagging algorithm
yields $bb$ or $\bar{b}\bar{b}$ we discard the events.  These detector
effects yield an additional $0.11$ dilution factor for the signal.

The gluino signal can be extracted using the basic acceptance cuts:
\begin{alignat}{7}
p_{T,b} &> 50 \gev  \qquad
&p_{T,\ell} &> 10\gev & \notag \\
p_{T,j}^{\rm min} &> 40 \gev
&p_{T,j}^{\rm max} &> 150 \gev & \notag \\
|\eta_i| &< 2.4 
&\Delta R_{ik} &> 0.4 & (i,k=b,j,\ell)
\label{eq:cuts1}
\end{alignat}
For the associated $\sq{}\go$ production we require the single jet from the
squark decay to pass the $p_{T,j}^{\rm max}$ cut. This selection of cuts
leaves us with $10 \fb$ of signal cross section from gluino pairs.  To reduce
the Standard Model backgrounds we apply the additional rejection cuts:
\begin{equation}
m_{\ell\ell}<80 \gev \quad
M_{\rm eff}>450\gev \quad
m_{jj}<300\gev\
\label{eq:cuts2}
\end{equation}
where $M_{\rm eff} = p_{T_{j,1}}+p_{T_{j,2}}+\sla{p}_T$. After this additional
cut our gluino--pair sample is $8.6 \fb$, with a $t\bar{t}jj$ background of
$34 \fb$. The associated $\sq{}\go$ production channel yields a rate ($85
\pb$) about ten times larger than the gluino pair sample while the Standard
Model $t\bar{t}j$ background to this channel is $23 \fb$ after cuts, which
means that both channels together range around $S/B \sim
1$~\cite{skands,madevent}. Our Standard Model and SUSY backgrounds originate
from wrongly combined and therefore uncorrelated leptons from independent
decays.  An efficient way to eliminate these backgrounds beyond the level $S/B
\sim 1$ is to subtract the measured opposite flavor dileptons from the same
flavor dileptons~\cite{Gjelsten:2004ki}. Because the precise prediction of the
remaining small backgrounds is beyond the scope of this paper we will not
include SUSY or Standard Model backgrounds in our analysis.\bigskip

\begin{figure}[t]
  \begin{center}
  \includegraphics[width=8.0cm]{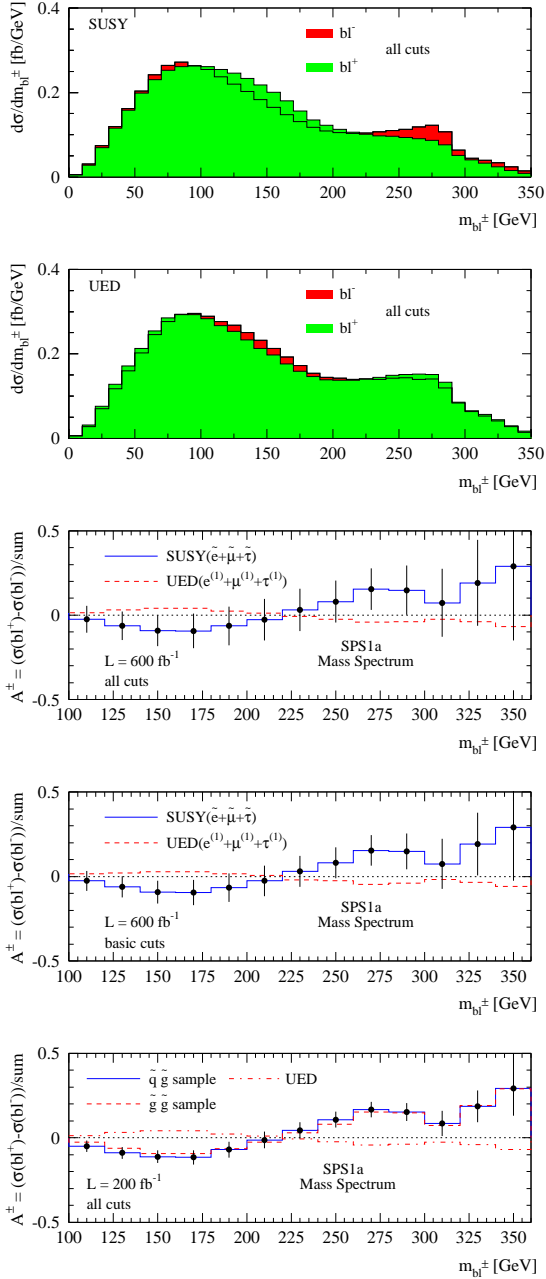}
  \end{center}
  \vspace*{-8mm}
  \caption{In the first panel we show the $b$--$\ell^\pm$ invariant mass
    distribution after cuts Eqs.~(\ref{eq:cuts1},\ref{eq:cuts2}) using only
    $\go\go$ production for the parameter point SPS1a.  The second panel shows
    the $m_{b\ell^\pm}$ spectrum for the UED interpretation assuming that the
    first KK states have masses equal to the SUSY particles in the first
    panel. The third panel contains the asymmetry $A^\pm(m_{b\ell})$ defined
    in Eq.~(\ref{eq:asymm}) after cuts Eqs.~(\ref{eq:cuts1},\ref{eq:cuts2})
    and for an integrated luminosity of $600 \ifb$. The fourth panel is the
    same, but after basic cuts Eq.~(\ref{eq:cuts1}) only.  The last panel
    shows $A^\pm(m_{b\ell})$ using $\go\go$ and $\sq{}\go$ production after
    all cuts and for an integrated luminosity of $200 \ifb$.  }
\label{fig:complete}
\end{figure}

In the two first panels of Fig.~\ref{fig:complete} we show the distributions
for the bottom--lepton invariant masses, both for the SUSY case and for the
UED cascade. To avoid using any information but the spin we assume the SPS1a
spectrum for the UED particles and normalize their production cross section
times branching fractions to the SUSY rate. Because we set the masses equal
for the two interpretations (to make the two scenarios indistinguishable in
the usual kinematic analysis of edges and thresholds) all additional
information in the shape of $m_{b\ell}$ should be equivalent to angular
correlations. The two mass distributions are similar, both for the two lepton
charges and for the SUSY vs.\ UED interpretations.  Notwithstanding, we can
construct a particularly sensitive asymmetry for each of the two
interpretations
\begin{equation}
  A^\pm(m_{b\ell}) = 
  \frac{d\sigma/dm_{b\ell^+} - d\sigma/dm_{b\ell^-}}
  {d\sigma/dm_{b\ell^+} + d\sigma/dm_{b\ell^-}}  
\label{eq:asymm}
\end{equation}
that is based on possibility of distinguishing $b$ from $\bar{b}$ through
their semi-leptonic decays.  This asymmetry is equivalent to an asymmetry in
$m_{b\ell^-}$ vs.\ $m_{\bar{b} \ell^-}$. Moreover, it has the advantage that
systematic uncertainties will cancel to a large (yet hard to specify) degree.
From the top two panels in Fig.~\ref{fig:complete} we see that the generally
most dangerous systematic error, namely the jet energy scale, will not impact
the distinction between a SUSY and an UED interpretations of the gluino
cascade decay: shifting the energy on the $x$ axes will, for small
$m_{b\ell}$, always enhance the asymmetry for one of the two interpretations
and reduce it for the other. We therefore concentrate on the certainly
dominant statistical errors in the binned distributions.\smallskip

The error bars for the asymmetry $A^\pm(m_{b\ell})$ shown in the third
panel of Fig.~\ref{fig:complete} correspond to the statistical error
per bin, assuming an integrated luminosity of $600 \ifb$ and taking
account only the $\go\go$ channel. Of course, an optimized measurement
of the asymmetry in SUSY and UED scenarios would rely on the shape
analysis to optimize the significance, but from
Fig.~\ref{fig:complete} it is obvious that for a hierarchical mass
spectrum it is no problem to distinguish a fermionic gluino from a
bosonic KK gluon from the (angular) correlations in their decay
chains. \smallskip

In the fourth panel of Fig.~\ref{fig:complete} we depict the
asymmetries after imposing only the acceptance cuts defined in
Eq.~(\ref{eq:cuts1}). We confirm that our results are not biased by the
harder background--rejection cuts. Finally, the last panel in
Fig.~\ref{fig:complete} shows the individual $\sq{}\go$ and $\go\go$
contributions for an integrated luminosity of $200 \ifb$.  Both
contributions can indeed be added naively, confirming our claim that
these distributions only carry information from angular correlations
in the decay kinematics.\smallskip

\begin{figure}[t]
  \begin{center}
  \includegraphics[width=8.5cm]{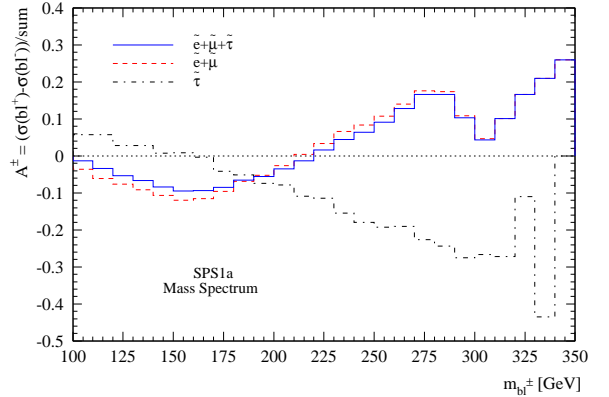}
  \end{center}
  \vspace*{-8mm}
\caption{Bottom--lepton asymmetry for the SUSY signal only. The 
  curves shown are for the first and second generation sleptons and
  for leptons coming from an intermediate $\stau{}$.}
\label{fig:staus}
\end{figure}

The details of the gluino decay chain reveal an important structure:
two leptons in the cascade decay usually come from an intermediate
first-- or second--generation slepton $\se{1,2}$, so we can use these
leptons to determine the $\se{1,2}$ masses from kinematical edges.
Alternatively, the cascade decay can proceed through a $\stau{1,2}$ with a
branching ratio of $6.3 \%$ as compared to $0.4 \%$ for the first--
and second generation sleptons combined.  Taking into account the
leptonic tau decays the branching fraction from $\stau{1,2}$ drops to
$0.2 \%$.\smallskip

\begin{figure}[t]
  \begin{center}
  \includegraphics[width=8.0cm]{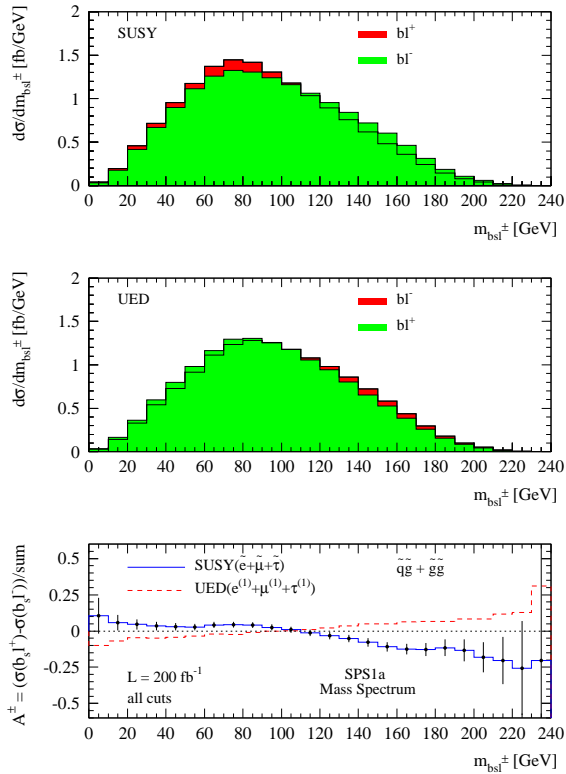}
  \end{center}
  \vspace*{-8mm}
  \caption{Softer bottom--lepton invariant mass distributions for SUSY (first
    panel) and UED (second panel) and asymmetry for the SPS1a mass spectrum
    (last panel), including the $\tau$ contribution. The mass distributions
    are shown adding the $\go\go$ and $\sq{}\go$ contributions after all
    cuts.}
\label{fig:asym_soft}
\end{figure}

For the parameter point SPS1a the (dominant) lighter selectron or smuon is
mostly right handed $\se{1} \sim \se{R}$, whereas the lighter stau is mostly
left handed $\stau{1} \sim \stau{L}$ due to the renormalization group running
and the fairly large $\tan\beta = 10$.  This means the contribution of the
stau to the mass asymmetry is opposite to the selectron and smuon
contributions.  In Fig.~\ref{fig:staus} we see how the $\stau{1}$ can in
principle wash out the asymmetry from selectrons and smuons. Luckily, the
impact of the $\stau{}$ on our asymmetry given in Eq.~(\ref{eq:asymm}) is
small because leptons from tau decays are softer and hence less likely pass
the cuts. After cuts the contribution from staus is about five times smaller
than the combined selectron and smuon signal.  We will further discuss the
different pattern for intermediate left and right handed sleptons as a general
feature for the gluino cascade in Sec.~\ref{sec:left_right}.\bigskip

As mentioned above, the SUSY spectrum might be such that it is possible to
identify the (near) bottom jet from the gluino decay since it is softer.  In
those cases where we can identify the softer $b$ jet with the near $b$ jet for
each event a similar asymmetry can be defined as
\begin{equation}
   A^\pm(m_{b_s\ell}) = 
       \frac{d\sigma/dm_{b_s\ell^+} - d\sigma/dm_{b_s\ell^-}}
            {d\sigma/dm_{b_s\ell^+} + d\sigma/dm_{b_s\ell^-}} 
\label{eq:asym_soft}
\end{equation}
Note that here the symbol $b$ means either $b$ or $\bar{b}$, without
distinction. Fig.~\ref{fig:asym_soft} shows that $A^\pm(m_{b_s\ell})$
can be an efficient tool to discriminate between SUSY and UED decay
cascades for a hierarchical mass spectrum.

\subsection{Purely Hadronic Correlations} 
\label{sec:had}

\begin{figure}[t]
  \begin{center}
  \includegraphics[width=6.5cm]{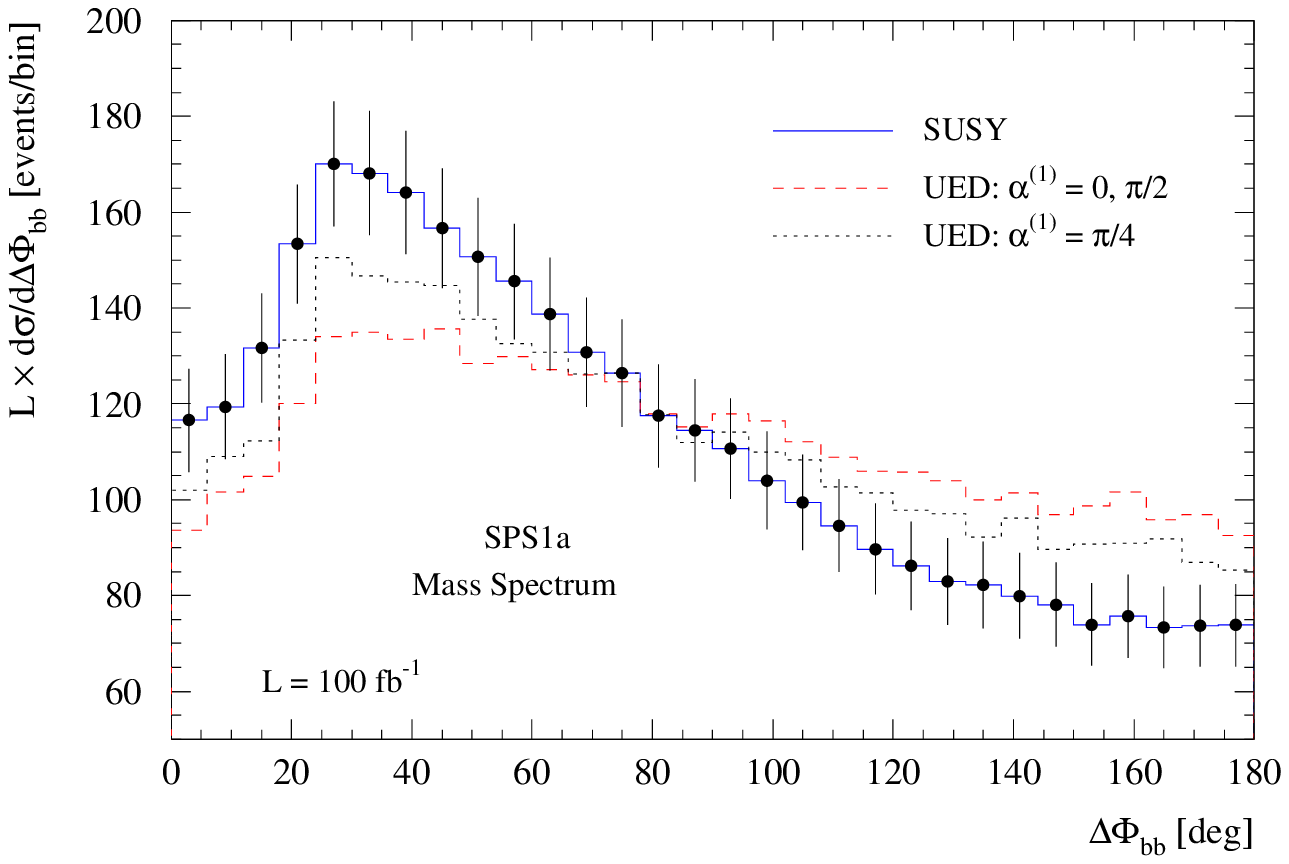} \\
  \includegraphics[width=6.5cm]{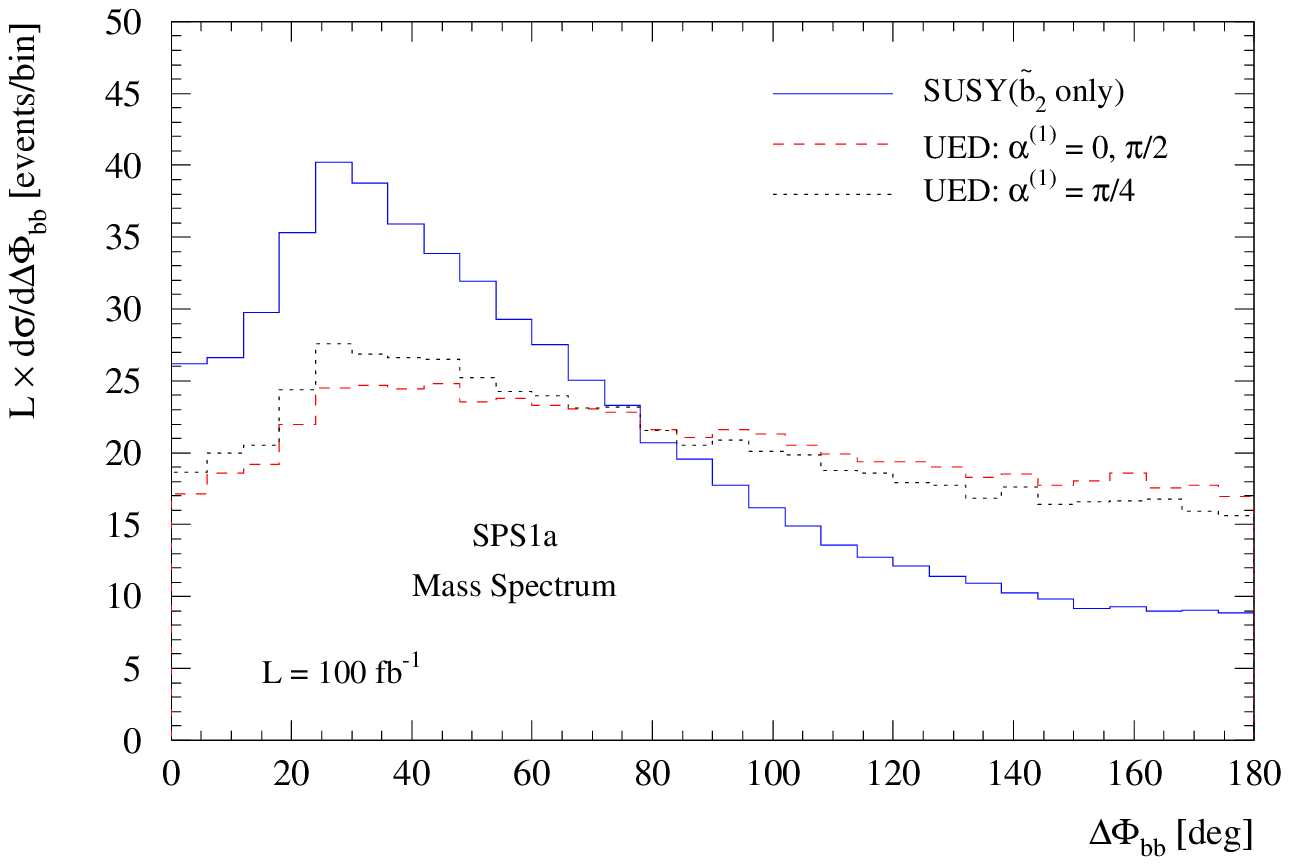} \\
  \includegraphics[width=6.5cm]{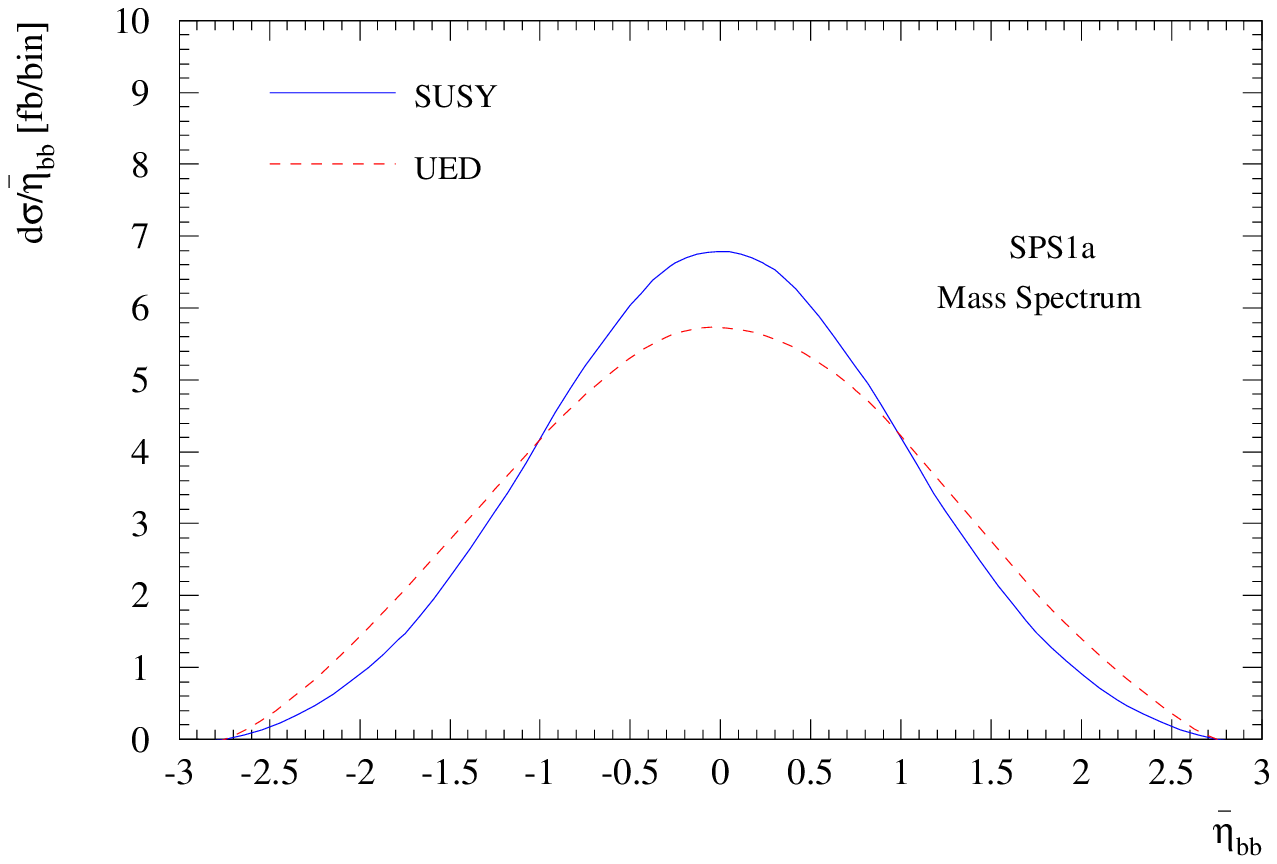} \\
  \end{center}
  \vspace*{-8mm}
  \caption{From the top: (i) azimuthal angle between the two bottom jets
    for UED and SUSY decay chains including 
    $\sbx{1}$ and $\sbx{2}$; (ii) including 
    only the effect of $\sbx{2}$ exchange; (iii) averaged
    bottom jet rapidities.  In all cases we assume the SPS1a spectrum 
    and add the rates for $\go \go$ and $\sq{} \go$ production.}
\label{fig:phibb}
\end{figure}

The correlation between a lepton and a bottom jet is only one of the
distributions we can use to distinguish the two interpretations of the
decay cascade. Unfortunately, it has been shown for squark decays that
purely leptonic distributions are not as useful as mixed lepton--jet
correlations~\cite{smillie}. However, in the gluino decay chain there
is an additional jet, so we can build purely hadronic correlations.
This has the advantage of being independent of the $\nn{2}$ decay,
which can involve not only intermediate sleptons, but also
intermediate gauge bosons or even three-body decay kinematics.\smallskip

In general, we expect all spin information to be hidden in angular
correlations. After exploiting the kinematic
endpoints to measure the masses in the cascade decays, we use the shape
of invariant mass distributions as a Lorentz--invariant formulation of
the angles. The only well defined angles we can observe at the LHC are
azimuthal angles between for example the two bottom jets, because they
are invariant under boosts in the beam direction. In
Fig.~\ref{fig:phibb} we present the distribution
$d\sigma/d\Delta\phi_{bb}$, which exhibits a distinct behavior for
SUSY and UED decay chains. These two possibilities can be disentangled
through the asymmetry:
 \begin{equation}
 \frac{\sigma(\Delta \Phi_{bb}< 90^o)-\sigma(\Delta \Phi_{bb}> 90^o)}
      {\sigma(\Delta \Phi_{bb}< 90^o)+\sigma(\Delta \Phi_{bb}> 90^o)}
\end{equation} 
This asymmetry assumes small values $0.08\pm 0.02$ for the UED spin assignment
with the usual mass--suppressed mixing angle $\alpha^{(1)} \sim 0$. On the
other hand, for the SUSY interpretation it is significantly larger $0.24\pm
0.02$. The quoted errors are statistical errors for the combination of the
gluino--pair and associated gluino--squark production channels and an
integrated luminosity of $100\ifb$. The UED cross section is as usually
normalized to the SUSY rate.\bigskip

To estimate the dependence of the $\Delta \Phi_{bb}$ distribution on
the couplings of the sbottom we present this distribution for a purely
$\sbx{2}$ decay chain in the second panel of Fig.~\ref{fig:phibb}.
Indeed, $\Delta \Phi_{bb}$ is insensitive to the left--right couplings
of bottom jets to the intermediate SUSY particles which makes it a
robust discriminating observable for spin correlations. This reflects 
the scalar nature of the intermediate sbottom. As a matter of fact, 
in analogy with the purely leptonic correlation~\cite{smillie} we find
that the different UED and SUSY behavior shown in Fig.~\ref{fig:phibb}
is mostly due to the boost of the heavy gluino or KK gluon.

According to Sec.~\ref{sec:ued} there is not very much room to modify the UED
Lagrangian to bring kinematical correlations closer to the SUSY prediction.
The KK weak mixing angle $\theta_w^{(n)}$ in Eq.~(\ref{eq:weinberg}) is fixed
by the interaction eigenstates' masses, so we can not change it while keeping
the masses fixed. The coupling structure in the decay matrix element is of the
general kind $(L^2+R^2)$, as long as the KK singlet and doublet fermions are
close in mass.  The same limitations hold when we try to adjust the mixing
between the singlet and doublet KK fermions, described by the angle
$\alpha^{(n)}$, Eq.~(\ref{eq:eigen}). In contrast to the 3rd--generation
sfermion sector in the MSSM, the UED mixing angle is not a (third) free
parameter, even if we move around the masses invoking boundary conditions.
Nevertheless, for illustration purpose we vary $\alpha^{(n)}$ in
Fig.~\ref{fig:phibb} to check whether the SUSY $\Delta \Phi_{bb}$ can be
reproduced by a UED decay chain with different couplings to the fermions.
From the two top panels of Fig.~\ref{fig:phibb} we see that the changes in the
UED distribution are not sufficient to mimic the SUSY predictions.\medskip

Our final observable is the average bottom rapidity~\cite{Meade:2006dw}
$\bar{\eta}_{bb}=(\eta_b + \eta_{\bar{b}})/2$ which we show in the
bottom panel of Fig.~\ref{fig:phibb}.  As we can see the bottom jets
from gluino cascades are typically more central than those from the
KK--gluon cascades, however, it is difficult to discriminate the SUSY
curve from UED on a bin-by-bin basis. Therefore, we define another
asymmetry
\begin{equation}
 \frac{\sigma(|\bar{\eta}_{bb}|< 1.0)-\sigma(|\bar{\eta}_{bb}|>1.0)}
      {\sigma(|\bar{\eta}_{bb}|< 1.0)+\sigma(|\bar{\eta}_{bb}|>1.0)} \; \; ,
\end{equation}
which gives $0.40\pm 0.02$ for SUSY and $0.24\pm 0.02$ for UED. These results
were obtained using the $\sq{} \go$ and $\go \go$ production channels and an
integrated luminosity of $100 \ifb$.  As always, we normalize the UED signal
to the SUSY rate.

\subsection{Degenerate UED-type Spectrum}
\label{sec:degenerate}

\begin{figure}[t]
  \begin{center}
  \includegraphics[width=8.5cm]{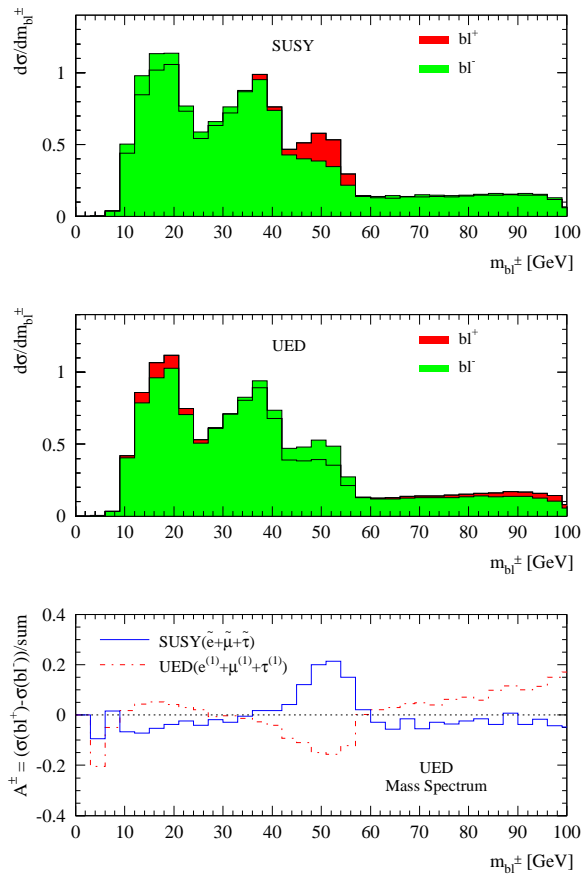}
  \end{center}
  \vspace*{-8mm}
  \caption{Bottom--lepton invariant mass distributions for SUSY (first panel)
    and UED (second panel) and asymmetry (third panel), in arbitrary
    units, for gluino pair
    production and after the background rejection cuts Eq.~(\ref{eq:cuts2}),
    for the UED mass spectrum described in Sec.~\ref{sec:ued}.}
\label{fig:ued}
\end{figure}

In the analysis above we have made a crucial assumptions: a hierarchical 
spectrum of the new particles responsible for the cascade decay.  In UED, the
first KK excitations will tend to be mass degenerate, unless this degeneracy
is broken by boundary conditions for the different fields or by large loop
corrections. For the highly degenerate spectrum listed in Sec.~\ref{sec:ued}
the outgoing fermions become very soft and the cross section after cuts
decreases, which translates into a strongly reduced precision of our
measurements.  Moreover, the invariant mass distributions shown in
Fig.~\ref{fig:ued} lose their characteristic pattern, for the SUSY as well as
for the UED prediction~\cite{smillie} and their associated asymmetries are
indistinguishable within the expected statistical errors. The same is
unfortunately true for the angular distributions of the $b$ jets. The hard
cuts imposed to separate the signal from $t\bar{t}$+jets backgrounds determine
completely the shape of angular distributions and invariant masses in both
descriptions.

\subsection{Left and Right Sleptons and Squarks}
\label{sec:left_right}

\begin{figure}[t]
  \begin{center}
  \includegraphics[width=8.5cm]{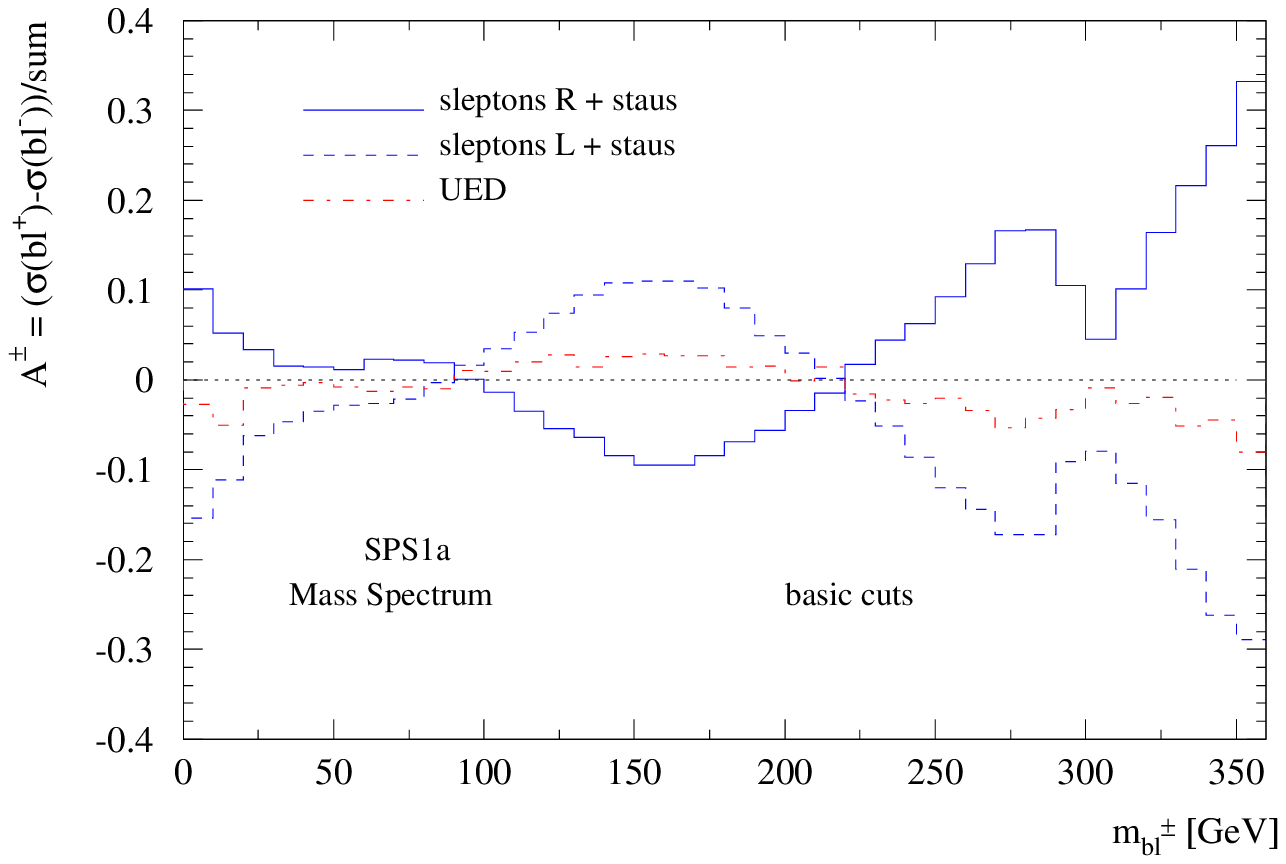} \\
  \includegraphics[width=8.5cm]{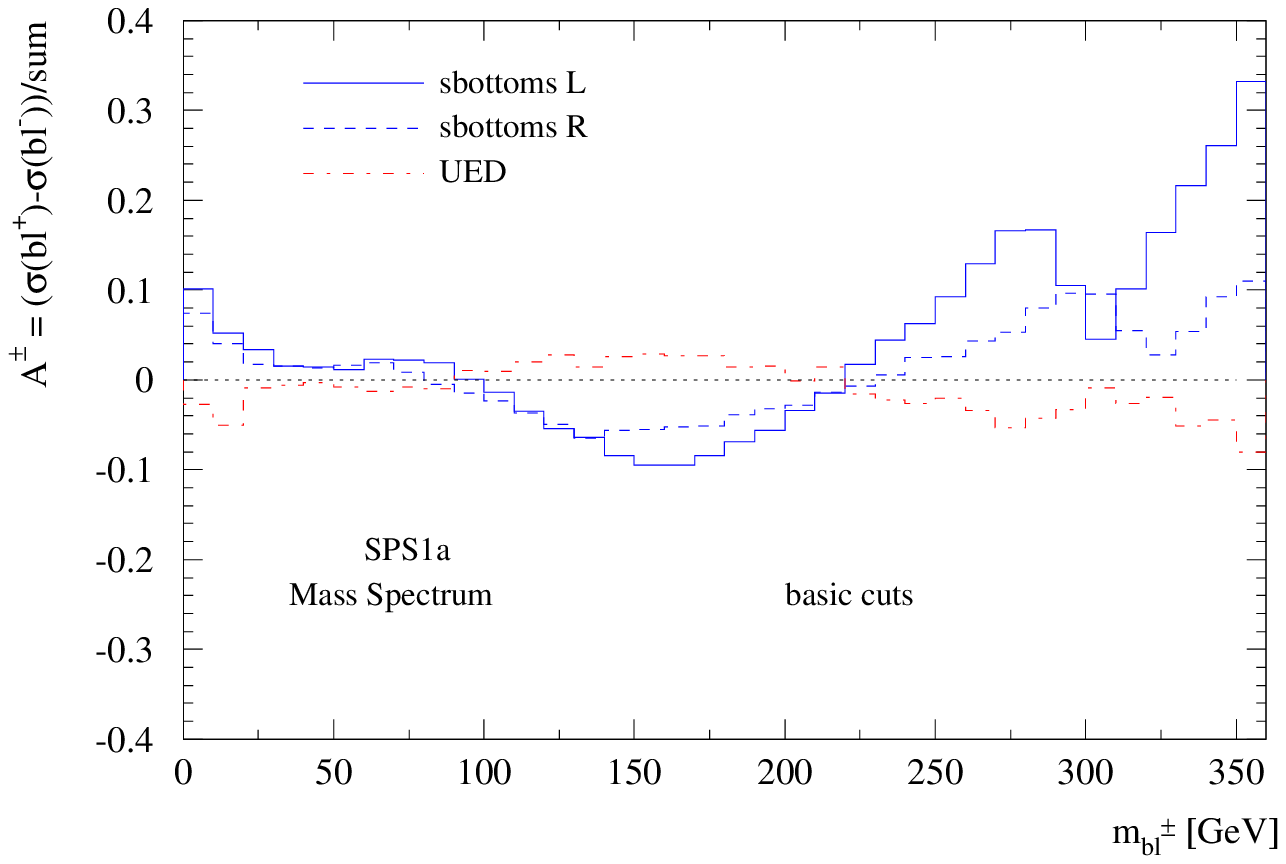} \\
  \end{center}
  \vspace*{-8mm}
  \caption{Asymmetry for an SPS1a mass spectrum and a supersymmetric gluino
    cascade, but varying the coupling of the cascade spectrum between purely
    left handed and purely right handed.}
\label{fig:left_right}
\end{figure}

As we point out in Sec.~\ref{sec:sps} the left handed and right handed
coupling of the slepton in the cascade is crucial to determine the asymmetry
in the lepton--bottom invariant mass. Or (in other words), the same
distributions we use to determine the spin of the cascade we can use to
determine the nature of the squark and slepton appearing in the cascade. This
twofold ambiguity is the major source of degeneracies in the determination of
the MSSM mass parameters at the LHC~\cite{sfitter,gordi}, and it can be broken
by the shape of $m_{b\ell}$ or by the variables $\Delta \phi_{b\bar{b}}$ and
$\bar{\eta}_{b\bar{b}}$.\smallskip

In the MSSM we are free to assign the two left and right soft--breaking
masses. For partners of essentially massless Standard Model particles the mass
eigenstates and the interaction eigenstates are identical. As mentioned above,
the light--flavor sleptons in the SPS1a parameter point are of the kind
$\se{1,2} \sim \se{R,L}$, the staus couple like $\stau{1,2} \sim \stau{L,R}$,
and the sbottoms like $\sbx{1,2} \sim \sbx{L,R}$. If we assume we know the
nature of the two lightest neutralinos we can roughly determine the nature of
a decaying squark from its branching fractions $\sq{} \to q \nn{1}$ and $\sq{}
\to q \nn{2} \to q \ell^+ \ell^- \nn{1}$, because the bino and the wino
fraction in the neutralino couple differently to left and right
sfermions.\smallskip

For the sleptons we usually cannot access branching fractions at the
LHC because we cannot rely on a direct production channel. For example
if the mass hierarchy is SPS1a-like ($\msl{2} > \mnn{2} > \msl{1}$)
squark and gluino cascade decays are the only source of information on
sleptons.  They are dominated by the lighter of the sleptons which is
produced on--shell in the cascade decay.  In that situation we can
determine the chiral structure of the slepton couplings from the same
distributions we use to distinguish a gluino cascade from a KK gluon
cascade. For the squark cascade this feature has been discussed
independent of the spin measurement~\cite{sleptons}. We illustrate
the link between slepton couplings and spins in the top panel of
Fig.~\ref{fig:left_right} where we display the asymmetry as a function
of $m_{b\ell^\pm}$ for left and right handed sleptons. The
asymmetry shows the opposite behavior for $\ell_R$ and $\ell_L$ and
consequently can be used as an indication of the $\se{1,2} \sim
\se{R,L}$ assignment. For scalar taus, Fig.~\ref{fig:staus} shows that
the same measurement is possible, provided we identify the tau leptons
from the cascade reliably~\cite{tautag}. On the other hand the
$\sbx{R}$ and $\sbx{L}$ contributions to the asymmetry are very
similar, as we can see in the bottom panel of
Fig.~\ref{fig:left_right}, so from these distributions we cannot
distinguish the two bottom states.

\subsection{Outlook} 

Proving the presence of a Majorana gluino is the prime task for the
LHC to show that new TeV--scale physics is supersymmetric. It has been
known for a long time that like--sign dileptons are a clear sign for
the Majorana nature of a newly found strongly interacting
particle~\cite{likesign}. The remaining loop hole in this argument is
to show that the gluino candidate is actually a fermion.  Recently, it
has been shown how to distinguish supersymmetric partners of Standard
Model particles from same-spin partners, for example described by UED
models~\cite{barr,smillie}.\smallskip

We extend these spin analyses to the case of a gluino decaying through
the bottom cascade. This is the decay chain which can best be used for
the gluino mass measurement~\cite{mgl}. The decay cascade can be
interpreted as a SUSY or as a UED signal, with identical particle
masses.  To distinguish the two spin patterns it is crucial to limit
the observables to angular correlations linked to the spins and to
ignore additional information which can come from production cross
sections times branching rations or from `typical' mass
spectra.\medskip

Using a list of asymmetries (constructed from lepton--bottom
correlations or from pure bottom--bottom correlations) we distinguish
between the SUSY and the UED cascade interpretations and thus
determine the spin of the gluino. The spin information which is
clearly present in the decay kinematics is always entangled with the
left and right handed sfermion couplings~\cite{sleptons}. Turning the
argument around, we find that the slepton coupling structure can be
determined from these kinds of correlations together with the spins.
This reduces possible degeneracies in the SUSY parameter
extraction~\cite{sfitter,gordi}.\bigskip

While spin analyses for SUSY models (using UED as a straw man) are an
exciting new development for the LHC, they are much more complex than
ILC spin analyses~\cite{tesla} because of the entanglement with the
left and right handed couplings of supersymmetric scalars. On the
other hand, the gluino will likely not be pair produced at the ILC.
For example, to test gaugino masses unification we need the gluino
spin measurement at the LHC to unambiguously identify the three
gauginos and evolve their masses to some high
scale~\cite{sfitter,pmz_unification}.

\begin{acknowledgments}
  We would like to thank Martin Schmaltz, Gustavo Burdman, Chris
  Lester, and Matthew Reece for insightful comments. TP would like to
  thank Fabio Maltoni and the CP3 in Louvain la Neuve, where this
  works was finalized, for their hospitality. This research was
  supported in part by Funda\c{c}\~{a}o de Amparo \`a Pesquisa do
  Estado de S\~ao Paulo (FAPESP) and by Conselho Nacional de
  Desenvolvimento Cient\'{\i}fico e Tecnol\'ogico (CNPq).
\end{acknowledgments}
 

\end{document}